\documentclass[conference]{IEEEtran}
\IEEEoverridecommandlockouts

\usepackage{cite}
\usepackage{amsmath,amssymb,amsfonts}
\usepackage{algorithmic}
\usepackage{graphicx}
\usepackage{textcomp}
\usepackage{xcolor}
\usepackage{float}   

\def\BibTeX{{\rm B\kern-.05em{\sc i\kern-.025em b}\kern-.08em
    T\kern-.1667em\lower.7ex\hbox{E}\kern-.125emX}}
\begin{document}

\title{LivePhys: Transforming Static Physics Problems into Interactive Simulations via a Scan-to-Play Framework
}

\author{
\IEEEauthorblockN{
Xiaowei Dai$^{*}$,
Ziyu Luo$^{*}$,
Xiangwen Zhang$^{*}$,
Xiaoming Chen (Corresponding Author)$^{*}$,
Juan Wu$^{\dagger}$,
Yonghong Ke$^{\ddagger}$%
}
\IEEEauthorblockA{
$^{*}$School of Computer and Artificial Intelligence, Beijing Technology and Business University, Beijing, China\\
(Emails: \{2330702014, 2431062101, 2330702053\}@st.btbu.edu.cn, xiaoming.chen@btbu.edu.cn) \\
$^{\dagger}$School of Educational Technology, Beijing Normal University, Beijing, China (Email: wuj@bnu.edu.cn)\\
$^{\ddagger}$School of Chinese Language and Literature, Beijing Normal University, Beijing, China (Email: yh8555@126.com)
}
}

\maketitle

\begin{abstract}
Physics problems in textbooks are typically presented as static diagrams accompanied by brief textual descriptions, requiring learners to infer dynamic physical behaviors through mental visualization. This process often imposes high cognitive demands and limits learners' ability to form accurate mental models. In this paper, we present \textbf{LivePhys}, a framework that enables a \emph{Scan-to-Play} paradigm for mechanics learning by transforming static textbook physics problems into executable, interactive simulations. LivePhys decouples multimodal perception from physics-aware reasoning and deterministic simulation. Given a problem diagram and its accompanying text, LivePhys performs text extraction, geometric segmentation, and cross-modal grounding to construct a structured, physics-aware intermediate representation. A multimodal large language model is then used as a reasoning controller to infer entities, parameters, and constraints, which are executed by a physics engine to generate spatially consistent and interactive simulations that allow learners to explore and manipulate problem conditions dynamically. Our evaluation results show that LivePhys significantly outperforms general-purpose multimodal models in simulation executability, spatial accuracy, and interaction fidelity. In addition, a user study demonstrates that interacting with LivePhys-generated simulations reduces learners' perceived cognitive load compared to static textbook materials. 
\end{abstract}


\section{INTRODUCTION}
In traditional physics education, mechanics problems are commonly presented as static diagrams in textbooks, accompanied by brief textual descriptions. In such settings, learners are expected to infer dynamic physical processes, such as motion, force, and interaction, from static visual and textual cues. This often requires learners to mentally simulate how a physical system evolves over time while repeatedly shifting attention between the diagram, the text, and their own mental visualization. As a result, many learners struggle to form accurate and coherent mental models, and the learning process imposes substantial cognitive demands, increasing intrinsic cognitive load \cite{sweller1988cognitive} and frequently inducing the split-attention effect \cite{kalyuga1999managing}.

Interactive simulations and virtual laboratories offer a promising alternative by externalizing physical dynamics into observable and manipulable environments, thereby supporting inquiry-based learning and conceptual understanding \cite{de1998scientific}. However, authoring high-quality physics simulations typically requires domain expertise and significant programming effort, which limits scalability across the large number of problems found in textbooks and educational materials. Recent advances in general-purpose multimodal large language models (MLLMs) have enabled the automatic generation of explanations or executable code from images and text, yet end-to-end generation remains unreliable for physics simulation tasks. In particular, the VideoPhy benchmark reveals significant physical commonsense violations in state-of-the-art text-to-video generative models \cite{bansal2024videophy}. Our task requires spatial layout consistency and physical plausibility, avoiding overlapping configurations, unstable dynamics, and incorrect interaction bindings. The central challenge, therefore, is how to reliably reconstruct an executable \emph{physics scenario} from static text and diagram inputs, one that exhibits geometrically consistent layouts, explicit physical constraints, and meaningful interactivity, without requiring manual authoring or expert intervention.

\begin{figure}[t]
    \centering
    \includegraphics[width=1\linewidth]{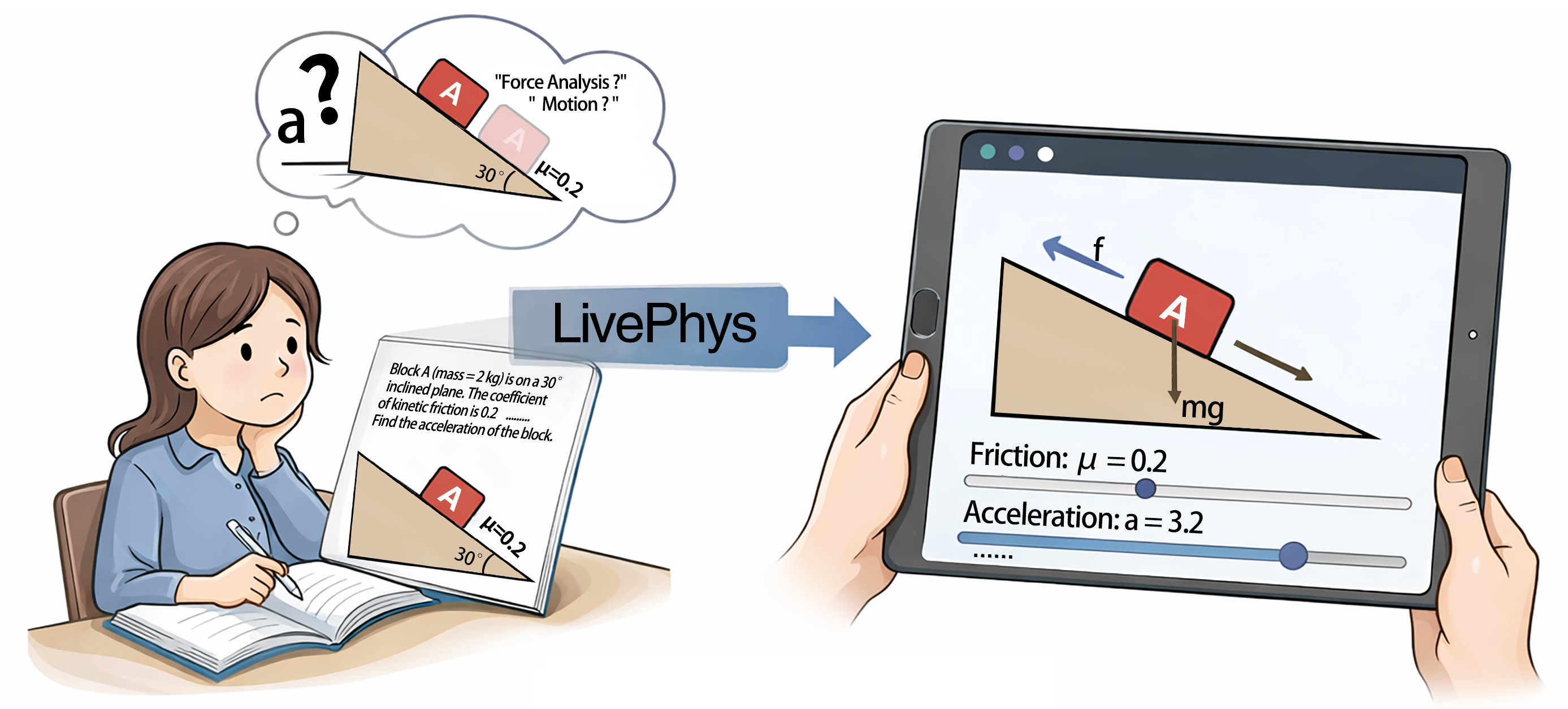}
    \caption{LivePhys transforms a physics problem (with a static diagram) into an interactive simulation with dynamic visualization that is geometrically consistent, enforces explicit physical constraints, and supports meaningful learner interaction.}
    \label{fig:fig0}
\end{figure}

To address this challenge, we propose \textbf{LivePhys}, a framework that enables a \emph{Scan-to-Play} paradigm for mechanics learning
(Fig.~\ref{fig:fig0}). LivePhys decouples multimodal perception from physics-aware reasoning and deterministic simulation: it first extracts and aligns textual and visual information into a structured intermediate representation (IR), then leverages an MLLM to infer physical parameters and constraints, and finally uses the ``Matter.js'' physics engine \cite{matterjs} to generate spatially consistent, interactive simulations.

\begin{figure*}[t]
\centering
\includegraphics[width=1\textwidth]{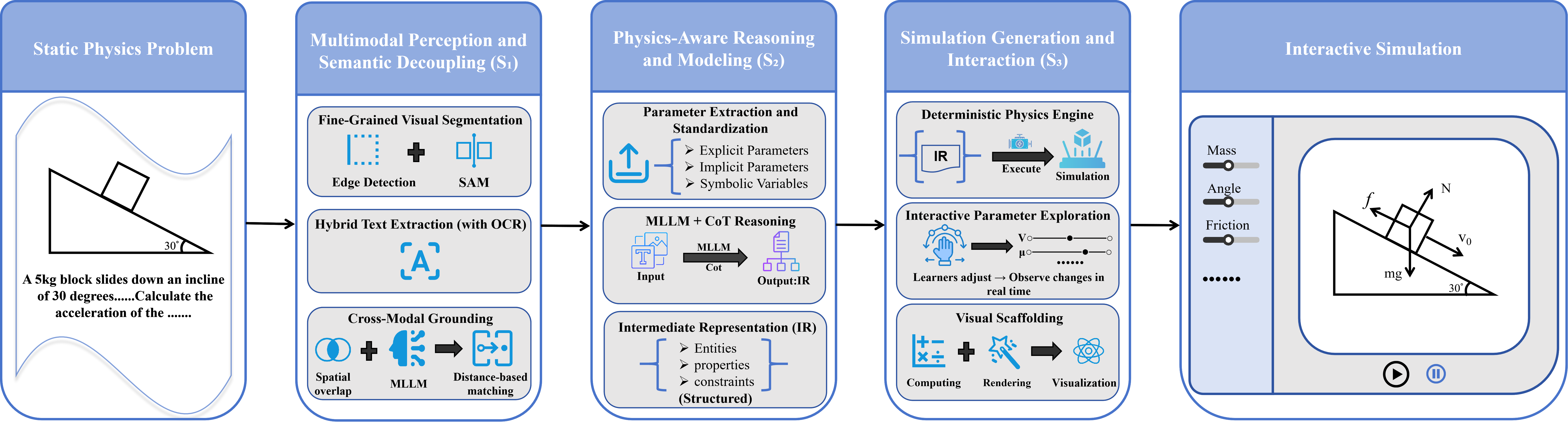}
\caption{Illustration of the architecture of LivePhys. LivePhys automates a \emph{Scan-to-Play} transformation through three cascading stages: (1) \emph{Multimodal Perception and Semantic Decoupling} ($S_1$), which employs optical character recognition (OCR) and visual segmentation to parse unstructured inputs and uses cross-modal grounding to align textual labels with visual entities in the textbook diagram; (2) \emph{Physics-Aware Reasoning and Modeling} ($S_2$), which leverages an MLLM to infer physical parameters and constraints and encapsulates them in a structured intermediate representation (IR); and (3) \emph{Simulation Generation and Interaction} ($S_3$), which transforms the intermediate representation into a deterministic physics simulation and provides learners with interactive controls for real-time parameter exploration.}
\label{fig:fig1_framework}
\end{figure*}

To summarize, this paper makes the following contributions:
\begin{itemize}
    \item We introduce \textbf{LivePhys}, a framework that transforms static textbook physics problems into executable and interactive simulations.
    \item We design and implement a three-stage generation pipeline that integrates a multimodal large language model with a physics engine, connecting multimodal perception to deterministic physics simulation.
\item We present both technical evaluations and a user study demonstrating that LivePhys improves simulation generation quality while reducing learners' perceived cognitive load.
\end{itemize}

\section{RELATED WORK}

\subsection{Interactive simulations and virtual laboratories}
 Interactive simulations can reduce cognitive load in STEM learning through dynamic visualizations \cite{moreno2007interactive}. In this area, for example, PhET \cite{perkins2006phet} offers research-based simulations with adjustable parameters.  Algodoo \cite{gregorcic2017algodoo} provides sandbox-style 2D physics experimentation. oPhysics \cite{ophysics} offers online physics simulations with predefined scenarios. Augmented Physics \cite{gunturu2024augmented} demonstrated automatic diagram-to-simulation conversion using multimodal techniques. However, these approaches require expert authoring or predefined scenarios, limiting scalability \cite{wieman2008phet}. To this end, our proposed LivePhys addresses this scalability barrier by automating the generation of problem-specific simulations.

\subsection{Multimodal models for scientific reasoning}
Recent multimodal models achieve strong performance on science benchmarks such as ScienceQA \cite{lu2022learn}. MLLMs, including GPT-4V \cite{achiam2023gpt} and LLaVA \cite{liu2023visual}, can align visual elements with textual constraints to produce answers and explanations. However, while they can describe dynamics in text, they generally do not provide end-to-end generation of interactive, physics-engine executable simulations. In contrast, our proposed LivePhys generates executable simulation environments directly from multimodal inputs.

\subsection{Generative video and physical grounding}
Text-to-video and diffusion-based video generation models can produce visually compelling motion \cite{liu2024sora}, including Video Diffusion Models \cite{ho2022video}. However, such models often lack reliable physical consistency, as they synthesize pixel-level dynamics statistically rather than enforcing deterministic physical laws, frequently resulting in physically implausible behavior. Existing work such as PhysGest \cite{dai2025physgest} generates videos from textbook diagrams but does not incorporate textual descriptions and relies on pre-rendered video streams. In contrast, LivePhys jointly reasons over both diagrams and text to generate executable simulations that support real-time parameter manipulation and inquiry-based exploration.

\section{METHODOLOGY}
\label{sec:methodology}

As illustrated in Fig.~\ref{fig:fig1_framework}, LivePhys transforms a textbook physics problem into an interactive web-based simulation through three consecutive stages: Multimodal Perception and Semantic Decoupling ($S_1$), Physics-Aware Reasoning and Modeling ($S_2$), and Simulation Generation and Interaction ($S_3$).

\subsection{Multimodal Perception and Semantic Decoupling ($S_1$)}
The objective of this stage is to parse the input textbook diagram $I$ into a discrete set of semantic objects. Given the specific nature of physics educational materials, LivePhys must process both geometric data and text containing mathematical notation. Here, \emph{semantic decoupling} means filtering out non-physical diagram primitives and retaining only physically meaningful entities aligned with text.

\subsubsection{Fine-grained Visual Segmentation}

To accurately extract physical entities (e.g., blocks, slopes) from the textbook diagram, we adopt a prompt-based segmentation strategy. We first apply edge detection~\cite{canny1986computational} to generate spatial prompts, which are then processed by the Segment Anything Model (SAM)~\cite{kirillov2023segment} to produce a set of candidate regions, each with an associated bounding box. A filtering step then removes non-physical elements such as dimension lines and arrows, retaining only objects that correspond to physical bodies in the problem.

\subsubsection{Hybrid Text Extraction}

Physics problems often interleave natural language with mathematical symbols (e.g., $v_0$, $\mu$). We employ a specialized optical character recognition (OCR) engine to extract a unified text stream $T$. For labels embedded within diagrams, the spatial positions of text elements are additionally recorded for subsequent cross-modal alignment.

\subsubsection{Cross-Modal Grounding}

A critical challenge is aligning textual references (e.g., ``Block A'') with their corresponding visual entities. We address this through a grounding mechanism that combines spatial overlap between text regions and object bounding boxes with distance-based matching. This process is further enhanced by the visual grounding capabilities of an MLLM, improving cross-modal alignment accuracy.

\subsection{Physics-Aware Reasoning and Modeling ($S_2$)}
This module serves as the core reasoning engine for generating the interactive simulation. We leverage an MLLM as the reasoning controller (e.g., GPT-4o in the current LivePhys implementation), while the LivePhys framework itself remains model-agnostic. Using Chain-of-Thought (CoT) prompting \cite{wei2022chain}, the MLLM identifies entities, infers constraints, and outputs a structured intermediate representation as detailed below.

\begin{figure}[H]
    \centering
    \includegraphics[width=1\linewidth]{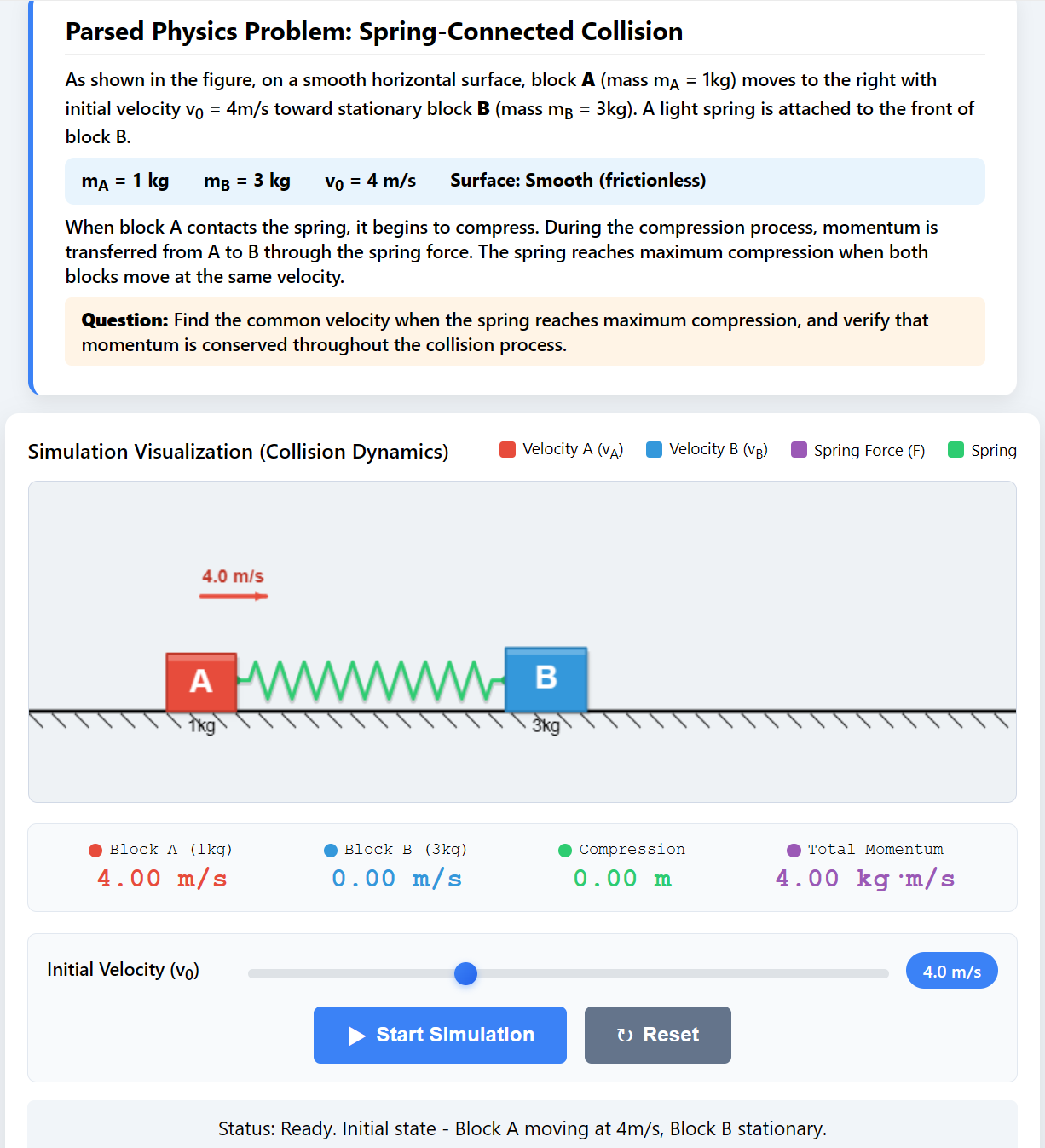}
    \caption{Interface of a complete generated simulation. It includes the parsed problem text (top), a spatially consistent simulation canvas for collision dynamics (middle), and interactive controls with state readouts (bottom), demonstrating LivePhys's end-to-end ``Scan-to-Play'' pipeline.}
    \label{fig:interface_showcase}
\end{figure}

  \begin{figure*}[t]
      \centering
      \includegraphics[width=1\textwidth]{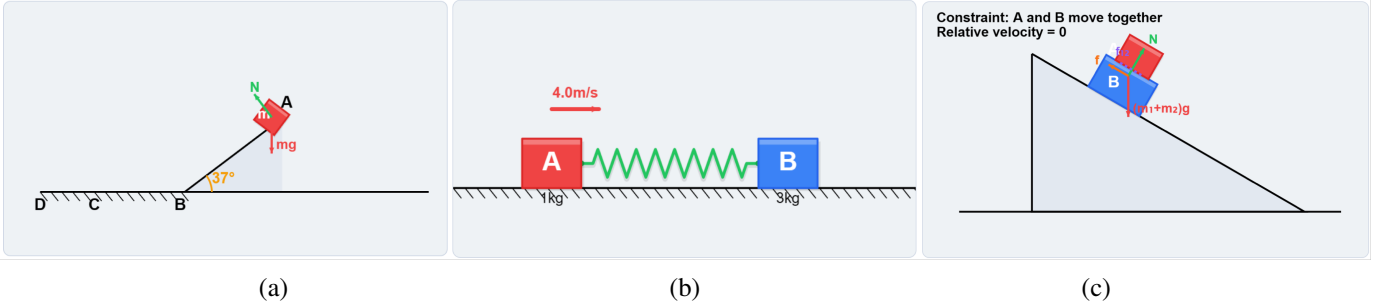}\\[2pt]
      \makebox[0.3\textwidth][c]{(a)}%
      \makebox[0.3\textwidth][c]{(b)}%
      \makebox[0.3\textwidth][c]{(c)}
      \caption{Examples of LivePhys-generated simulations with visual scaffolding across different mechanics scenarios:
  (a) incline dynamics with force labels; (b) spring-connected collision with velocity annotation and motion indicator; (c) stacked blocks on an incline with contact and constraint cues.}
      \label{fig:qualitative_comparison}
  \end{figure*}
  
\subsubsection{Parameter Extraction and Standardization}
The MLLM is instructed to identify and categorize three types of parameters:
\begin{itemize}
    \item Explicit Parameters: Numerical values directly extracted from the text (e.g., ``mass of 2kg'').
    \item Implicit Parameters: Values derived from domain knowledge mapping. LivePhys then embeds a set of domain-specific mapping rules within the MLLM prompt, e.g., ``\textit{smooth surface}'' $\rightarrow \mu = 0$, ``\textit{free fall}'' $\rightarrow a = g = 9.8\,\text{m/s}^2$, etc.
    \item Symbolic Variables: Variables without specific values (e.g., ``length $L$''). LivePhys preserves them as symbolic tokens to be exposed for user interaction.
\end{itemize}

\subsubsection{Intermediate Representation (IR)}

All reasoning outputs are encoded in a structured intermediate representation that explicitly captures physical entities, their properties (e.g., mass, position, shape), and inter-body relationships such as contact and support constraints. This structured format ensures that all information required for deterministic simulation is preserved.

\subsection{Simulation Generation and Interaction ($S_3$)}
  The final stage transforms the intermediate representation into an interactive physics simulation. We employ ``Matter.js'', a deterministic physics engine, which executes the structured scene specification to produce physically plausible behavior. The rendering layer maps simulation coordinates to screen space based on the detected object dimensions, preserving spatial consistency with the original diagram. Fig.~\ref{fig:interface_showcase} shows an illustration of the simulation interface.                     

\subsubsection{Interactive Parameter Exploration}
When a problem contains symbolic variables, e.g., unspecified initial velocity $v_0$ or friction coefficient $\mu$, LivePhys automatically generates interactive controls that allow learners to manipulate these parameters and observe the resulting changes in real time. For example, a learner can adjust the initial velocity of an object and immediately observe how the trajectory and collision dynamics change accordingly. This supports inquiry-based learning~\cite{de1998scientific} by enabling learners to formulate and test hypotheses about cause-effect relationships in mechanics.

\subsubsection{Visual Scaffolding}

To bridge abstract physics concepts and observable phenomena, LivePhys can augment the simulation with dynamic visual overlays, such as velocity annotations, motion indicators, force labels, as demonstrated in Fig.~\ref{fig:qualitative_comparison}. These scaffolds help learners connect symbolic representations to their visual manifestations, supporting the formation of coherent mental models.

\section{EVALUATION RESULTS}

\subsection{Textbook Problem Collection}
We collected 50 representative mechanics problems (containing rigid-body objects) from commonly used physics textbooks and educational materials (all textbook problems are used solely for non-commercial research and educational purposes). Each problem consists of a static diagram paired with a textual description, reflecting the standard format encountered in classroom instruction and assessments. The selected problems cover common diagrammatic structures such as stacked blocks, carts, inclined planes, and collision or contact-based interactions, all of which require accurate spatial interpretation for stable rigid-body simulation.

\begin{table}[h]
\centering
\caption{Comparison of simulation generation quality across methods.}
\label{tab:llm_comparison}
\resizebox{\columnwidth}{!}{
  \begin{tabular}{lccc}
  \hline
  \textbf{Method} & \textbf{Executability Rate} & \textbf{Spatial Accuracy} & \textbf{Interaction Fidelity} \\
  \hline
  GPT-4o & 78.0\% & 46.0\% & 60.0\% \\
  Claude 3.5 Sonnet & 82.0\% & 48.0\% & 64.0\% \\
  Gemini 1.5 Pro & 76.0\% & 42.0\% & 58.0\% \\
  \textbf{LivePhys (Ours)} & \textbf{94.0\%} & \textbf{92.0\%} & \textbf{90.0\%} \\
  \hline
  \end{tabular}
}
\end{table}

\subsection{Evaluation on Generated Simulation Quality}
We compare LivePhys against three widely used general-purpose MLLMs that educators may plausibly attempt to use for simulation generation: GPT-4o (OpenAI) \cite{hurst2024gpt}, Claude 3.5 Sonnet (Anthropic) \cite{anthropic2024claude35}, and Gemini 1.5 Pro (Google) \cite{team2024gemini}. To simulate a realistic workflow, all baselines were evaluated using a standardized zero-shot prompt: ``\textit{Write a complete Matter.js simulation code based on this image. Ensure the layout matches the diagram and include interactive sliders.}'' In contrast, LivePhys follows its designed Understand-then-Simulate pipeline.

Following the execution-based evaluation methodology for code generation~\cite{chen2021evaluating}, we define three task-specific metrics:
 \begin{itemize}
      \item Executability Rate (ER): the percentage of generated simulations that execute without syntax errors or runtime failures.
      \item Spatial Accuracy (SA): the percentage of simulations whose initial spatial layout matches the source diagram without overlaps or misplacements.
      \item Interaction Fidelity (IF): the percentage of simulations in which interactive user interface controls are correctly bound to physical parameters.
  \end{itemize}
ER was verified by automated execution; SA and IF were scored as binary pass/fail by two authors independently, with disagreements resolved by discussion.

As shown in Table~\ref{tab:llm_comparison}, LivePhys consistently outperforms all baselines. While baseline models often generate syntactically valid code, they struggle to infer precise spatial layouts from diagrams, leading to overlapping objects or unstable configurations. LivePhys mitigates these issues through fine-grained visual segmentation and explicit coordinate mapping.


\subsection{Evaluation on Cognitive Load during Learning}
To assess the pedagogical effects of LivePhys, we conducted a user study involving 20 undergraduate students aged 18--24 ($M = 20.5$, $SD = 1.8$), who provided informed consent prior to participation. Participants were randomly assigned to a Control Group ($n=10$), which used textbook problems, or a LivePhys Group ($n=10$), which interacted with LivePhys-generated simulations. Perceived cognitive load was measured using the NASA Task Load Index (NASA-TLX) \cite{hart1988development}. We focused on three dimensions most relevant to cognitive processing in problem-solving tasks: Mental Demand, Effort, and Frustration. Group differences were analyzed using independent samples $t$-tests. Both groups received the same set of problems and the same time budget.

\begin{figure}[t]
    \centering
    \includegraphics[width=1\linewidth]{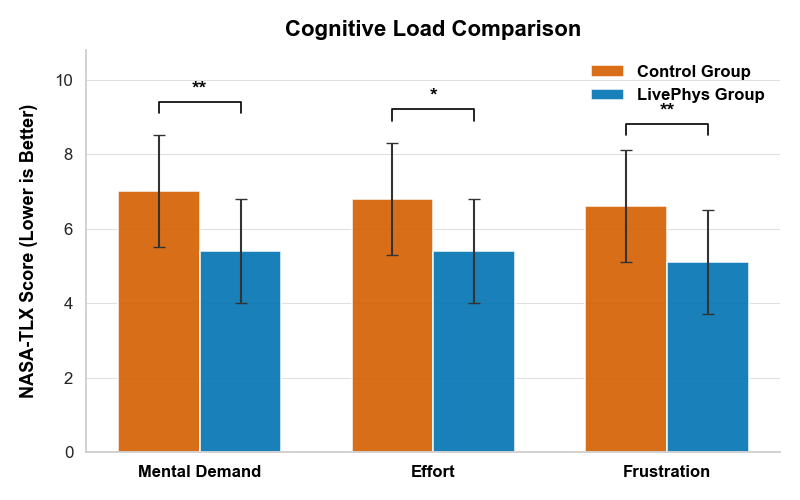}
  \caption{Comparison of cognitive load based on NASA-TLX scores.
  Participants in the LivePhys Group reported significantly lower Mental
  Demand, Effort, and Frustration than those in the Control Group
  (** $p<.01$, * $p<.05$).}
    \label{fig:user_study}
\end{figure}

As shown in Fig.~\ref{fig:user_study}, participants in the LivePhys Group reported significantly lower scores across all three NASA-TLX dimensions: Mental Demand ($p<.01$, $d=1.52$), Effort ($p<.05$, $d=1.10$), and Frustration ($p<.01$, $d=1.50$). Given the limited sample size, these results provide preliminary evidence suggesting that interactive simulations may reduce learners' perceived cognitive load compared to static materials.

\section{CONCLUSIONS AND DISCUSSION}
\label{sec:conclusion}
  We presented LivePhys, a framework that enables a \emph{Scan-to-Play} paradigm by transforming textbook physics problems with static diagrams into physically grounded, interactive simulations. By decoupling visual perception from symbolic reasoning, LivePhys mitigates spatial layout inconsistencies and bridges the gap between diagram understanding and inquiry-based physics learning. Our evaluation demonstrates that LivePhys substantially outperforms general-purpose multimodal language models in generating physically consistent and executable simulations. A user study provides preliminary evidence that interactive simulations can reduce learners' perceived cognitive load compared to traditional static learning materials. However, LivePhys has several limitations. The user study involved a relatively small sample and measured only perceived workload via NASA-TLX, which does not directly map onto the tripartite cognitive load framework in educational psychology. Future work should incorporate larger samples, direct learning outcome measures, and more fine-grained cognitive load instruments to validate the pedagogical effectiveness of LivePhys. While the current implementation of LivePhys focuses on 2D mechanics, future work will also extend the framework toward richer object representations and more immersive learning experiences, including 3D scene understanding and virtual reality--based educational interfaces.

\section*{ACKNOWLEDGMENT}
This work was supported in part by the National Natural Science Foundation of China (NSFC) under Grant 62577004 and the Fundamental Research Funds for the Central Universities under Grant 1253200047.






\bibliographystyle{IEEEtran}
\bibliography{refs} 

@inproceedings{gunturu2024augmented,
  title={Augmented Physics: Creating interactive and embedded physics simulations from static textbook diagrams},
  author={Gunturu, Aditya and Wen, Yi and Zhang, Nandi and Thundathil, Jarin and Kazi, Rubaiat Habib and Suzuki, Ryo},
  booktitle={Proceedings of the 37th Annual ACM Symposium on User Interface Software and Technology},
  pages={1--12},
  year={2024}
}

@article{perkins2006phet,
  title={PhET: Interactive simulations for teaching and learning physics},
  author={Perkins, Katherine and Adams, Wendy and Dubson, Michael and Finkelstein, Noah and Reid, Sam and Wieman, Carl and LeMaster, Ron},
  journal={The physics teacher},
  volume={44},
  number={1},
  pages={18--23},
  year={2006},
  publisher={American Association of Physics Teachers}
}

@article{gregorcic2017algodoo,
  title={Algodoo: A tool for encouraging creativity in physics teaching and learning},
  author={Gregorcic, Bor and Bodin, Madelen},
  journal={The Physics Teacher},
  volume={55},
  number={1},
  pages={25--28},
  year={2017},
  publisher={AIP Publishing}
}

@misc{ophysics,
  author       = {Tom Walsh},
  title        = {oPhysics},
  year         = {2025},
  howpublished = {\url{https://ophysics.com/}},
  note         = {Accessed: August 23, 2025}
}

@article{sweller1988cognitive,
  title={Cognitive load during problem solving: Effects on learning},
  author={Sweller, John},
  journal={Cognitive Science},
  volume={12},
  number={2},
  pages={257--285},
  year={1988},
  publisher={Elsevier}
}

@article{kalyuga1999managing,
  title={Managing split-attention and redundancy in multimedia instruction},
  author={Kalyuga, Slava and Chandler, Paul and Sweller, John},
  journal={Applied Cognitive Psychology: The Official Journal of the Society for Applied Research in Memory and Cognition},
  volume={13},
  number={4},
  pages={351--371},
  year={1999},
  publisher={Wiley Online Library}
}

@article{de1998scientific,
  title={Scientific discovery learning with computer simulations of conceptual domains},
  author={De Jong, Ton and Van Joolingen, Wouter R},
  journal={Review of educational research},
  volume={68},
  number={2},
  pages={179--201},
  year={1998},
  publisher={Sage Publications Sage CA: Thousand Oaks, CA}
}

@article{wieman2008phet,
  title={PhET: Simulations that enhance learning},
  author={Wieman, Carl E and Adams, Wendy K and Perkins, Katherine K},
  journal={Science},
  volume={322},
  number={5902},
  pages={682--683},
  year={2008},
  publisher={American Association for the Advancement of Science}
}

@article{ho2022video,
  title={Video diffusion models},
  author={Ho, Jonathan and Salimans, Tim and Gritsenko, Alexey and Chan, William and Norouzi, Mohammad and Fleet, David J},
  journal={Advances in neural information processing systems},
  volume={35},
  pages={8633--8646},
  year={2022}
}

@article{moreno2007interactive,
  title={Interactive multimodal learning environments: Special issue on interactive learning environments: Contemporary issues and trends},
  author={Moreno, Roxana and Mayer, Richard},
  journal={Educational psychology review},
  volume={19},
  number={3},
  pages={309--326},
  year={2007},
  publisher={Springer}
}

@article{lu2022learn,
  title={Learn to explain: Multimodal reasoning via thought chains for science question answering},
  author={Lu, Pan and Mishra, Swaroop and Xia, Tanglin and Qiu, Liang and Chang, Kai-Wei and Zhu, Song-Chun and Tafjord, Oyvind and Clark, Peter and Kalyan, Ashwin},
  journal={Advances in Neural Information Processing Systems},
  volume={35},
  pages={2507--2521},
  year={2022}
}

@article{bansal2024videophy,
  title={Videophy: Evaluating physical commonsense for video generation},
  author={Bansal, Hritik and Lin, Zongyu and Xie, Tianyi and Zong, Zeshun and Yarom, Michal and Bitton, Yonatan and Jiang, Chenfanfu and Sun, Yizhou and Chang, Kai-Wei and Grover, Aditya},
  journal={arXiv preprint arXiv:2406.03520},
  year={2024}
}

@article{liu2024sora,
  title={Sora: A review on background, technology, limitations, and opportunities of large vision models},
  author={Liu, Yixin and Zhang, Kai and Li, Yuan and Yan, Zhiling and Gao, Chujie and Chen, Ruoxi and Yuan, Zhengqing and Huang, Yue and Sun, Hanchi and Gao, Jianfeng and others},
  journal={arXiv preprint arXiv:2402.17177},
  year={2024}
}

@inproceedings{kirillov2023segment,
  title={Segment anything},
  author={Kirillov, Alexander and Mintun, Eric and Ravi, Nikhila and Mao, Hanzi and Rolland, Chloe and Gustafson, Laura and Xiao, Tete and Whitehead, Spencer and Berg, Alexander C and Lo, Wan-Yen and others},
  booktitle={Proceedings of the IEEE/CVF international conference on computer vision},
  pages={4015--4026},
  year={2023}
}

@article{wei2022chain,
  title={Chain-of-thought prompting elicits reasoning in large language models},
  author={Wei, Jason and Wang, Xuezhi and Schuurmans, Dale and Bosma, Maarten and Xia, Fei and Chi, Ed and Le, Quoc V and Zhou, Denny and others},
  journal={Advances in neural information processing systems},
  volume={35},
  pages={24824--24837},
  year={2022}
}

@incollection{hart1988development,
  title={Development of NASA-TLX (Task Load Index): Results of empirical and theoretical research},
  author={Hart, Sandra G and Staveland, Lowell E},
  booktitle={Advances in psychology},
  volume={52},
  pages={139--183},
  year={1988},
  publisher={Elsevier}
}

@misc{hurst2024gpt,
  author       = {{OpenAI}},
  title        = {{GPT-4o}},
  year         = {2024},
  howpublished = {\url{https://openai.com/index/gpt-4o-system-card/}},
  note         = {Accessed: Apr. 30, 2026}
}

@misc{anthropic2024claude35,
  author       = {{Anthropic}},
  title        = {{Claude 3.5 Sonnet}},
  year         = {2024},
  howpublished = {\url{https://www.anthropic.com/news/claude-3-5-sonnet}},
  note         = {Accessed: Apr. 30, 2026}
}

@misc{team2024gemini,
  author       = {{Google}},
  title        = {{Gemini 1.5 Pro}},
  year         = {2024},
  howpublished = {\url{https://gemini.google.com/app}},
  note         = {Accessed: Apr. 30, 2026}
}

@misc{matterjs,
  title = {Matter.js: A 2D rigid body physics engine for the web},
  author = {Liam Brummitt},
  year = {2014},
  howpublished = {\url{https://github.com/liabru/matter-js}},
}

@article{canny1986computational,
  title={A computational approach to edge detection},
  author={Canny, John},
  journal={IEEE Transactions on pattern analysis and machine intelligence},
  number={6},
  pages={679--698},
  year={1986},
  publisher={Ieee}
}

@inproceedings{dai2025physgest,
  title={PhysGest: Transforming Static Textbook Diagrams into Physically Realistic Videos for Enhanced Physics Education},
  author={Dai, Xiaowei and Zhang, Xiangwen and Zhang, Qian and Hu, Zeke Zexi and Qian, Baotong and Chen, Xiaoming},
  booktitle={SIGGRAPH Asia 2025 Educator's Forum},
  pages={1--6},
  year={2025}
}

@article{liu2023visual,
  title={Visual instruction tuning},
  author={Liu, Haotian and Li, Chunyuan and Wu, Qingyang and Lee, Yong Jae},
  journal={Advances in neural information processing systems},
  volume={36},
  pages={34892--34916},
  year={2023}
}

@article{achiam2023gpt,
  title={Gpt-4 technical report},
  author={Achiam, Josh and Adler, Steven and Agarwal, Sandhini and Ahmad, Lama and Akkaya, Ilge and Aleman, Florencia Leoni and Almeida, Diogo and Altenschmidt, Janko and Altman, Sam and Anadkat, Shyamal and others},
  journal={arXiv preprint arXiv:2303.08774},
  year={2023}
}

@article{chen2021evaluating,
  title={Evaluating large language models trained on code},
  author={Chen, Mark and Tworek, Jerry and Jun, Heewoo and Yuan, Qiming and Pinto, Henrique Ponde De Oliveira and Kaplan, Jared and Edwards, Harri and Burda, Yuri and Joseph, Nicholas and Brockman, Greg and others},
  journal={arXiv preprint arXiv:2107.03374},
  year={2021}
}




\end{document}